\documentclass[10pt,prd,twocolumn, nofootinbib,preprint,superscriptaddress]{revtex4}
\pdfoutput=1
\usepackage[T1]{fontenc}
\usepackage{amsmath,amssymb}
\usepackage{epsfig}
\usepackage{float}
\usepackage{graphicx}
\usepackage[usenames,dvipsnames]{color}
\usepackage{subfigure}
\usepackage{slashed}
\usepackage[colorlinks,citecolor=blue]{hyperref}
\usepackage{pdfpages}
\usepackage{color}

\begin{document}
	\title{Type II Dirac Seesaw with Observable $\Delta N_{\rm eff}$ in the light of W-mass Anomaly}

	\author{Debasish Borah}
	\email{dborah@iitg.ac.in}
	\affiliation{Department of Physics, Indian Institute of Technology Guwahati, Assam 781039, India}
	
	\author{Satyabrata Mahapatra}
	\email{ph18resch11001@iith.ac.in}
	\affiliation{Department of Physics, Indian Institute of Technology Hyderabad, Kandi, Sangareddy 502285, Telangana, India}
	
	\author{Dibyendu Nanda}
	\email{dnanda@kias.re.kr}
\affiliation{School of Physics, Korea Institute for Advanced Study, Seoul 02455, Korea}

	\author{Narendra Sahu}
	\email{nsahu@phy.iith.ac.in}
	\affiliation{Department of Physics, Indian Institute of Technology Hyderabad, Kandi, Sangareddy 502285, Telangana, India}
	
	\begin{abstract}
	We propose a type II seesaw model for light Dirac neutrinos to provide an explanation for the recently reported anomaly in W boson mass by the CDF collaboration with $7\sigma$ statistical significance. In the minimal model, the required enhancement in W boson mass is obtained at tree level due to the vacuum expectation value of a real scalar triplet, which also plays a role in generating light Dirac neutrino mass. Depending upon the couplings and masses of newly introduced particles, we can have thermally or non-thermally generated relativistic degrees of freedom $\Delta N_{\rm eff}$ in the form of right handed neutrinos which can be observed at future cosmology experiments. Extending the model to a radiative Dirac seesaw scenario can also accommodate dark matter and lepton anomalous magnetic moment.
	\end{abstract}
	\maketitle
	\noindent

\noindent
\textbf{\emph{Introduction}:}  The CDF collaboration has provided an updated measurement of the W boson mass $M_W = 80433.5 \pm 9.4$ MeV \cite{CDF:2022hxs} using the data corresponding to 8.8 ${\rm fb}^{-1}$ integrated luminosity collected at the CDF-II detector of Fermilab Tevatron collider. This newly measured value has 7$\sigma$ deviation from the standard model (SM) expectation ($M_{W}=80357\pm6$ MeV). This has led to several discussions on possible implications and interpretations in the last week related to effective field theory \cite{Fan:2022yly, Bagnaschi:2022whn}, electroweak precision parameters \cite{deBlas:2022hdk, Strumia:2022qkt, Asadi:2022xiy, Lu:2022bgw}, beyond standard model (BSM) physics like dark matter (DM) \cite{Fan:2022dck, Zhu:2022tpr, Zhu:2022scj,Kawamura:2022uft, Nagao:2022oin, Liu:2022jdq}, additional scalar fields \cite{Sakurai:2022hwh, Cacciapaglia:2022xih,Song:2022xts, Bahl:2022xzi,Cheng:2022jyi,Babu:2022pdn,Heo:2022dey,Ahn:2022xeq,Zheng:2022irz, Perez:2022uil,Kanemura:2022ahw}, supersymmetry \cite{Du:2022pbp, Tang:2022pxh,Yang:2022gvz, Athron:2022isz,Ghoshal:2022vzo} and several others \cite{Yuan:2022cpw, Athron:2022qpo, Blennow:2022yfm, Heckman:2022the,Lee:2022nqz, DiLuzio:2022xns,Paul:2022dds,Biekotter:2022abc,Balkin:2022glu,Cheung:2022zsb,Du:2022brr, Endo:2022kiw, Crivellin:2022fdf, Arias-Aragon:2022ats}. Assuming this anomaly to be originating from beyond standard model (BSM) physics, here we consider a seesaw model for Dirac neutrinos where a real scalar triplet plays a non-trivial role in generating the required enhancement in W boson mass as well as light neutrino masses. Due to the existence of additional light species in the form of right chiral parts of light Dirac neutrinos, we can get enhancement in effective relativistic degrees of freedom $\Delta N_{\rm eff}$ depending upon Yukawa couplings and masses of additional particles including those involving the triplet scalar. We show that such enhanced $\Delta N_{\rm eff}$ can not only be constrained by existing data from the Planck collaboration, but also remains within reach of next generation cosmology experiments. After discussing the minimal model of tree level Dirac seesaw, we consider a radiative version of it which can accommodate dark matter as well as lepton anomalous magnetic moments, which are also signatures of BSM physics. 
\vspace{0.2cm}

\noindent
\textbf{\emph{Tree level Dirac seesaw}:} While most of the neutrino mass models invoke the Majorana nature of neutrinos, the possibility of Dirac seesaw is relatively less explored yet equally appealing. Different BSM frameworks for origin of light Dirac neutrino mass can be found in \cite{Babu:1988yq, Peltoniemi:1992ss, Farzan:2012sa, Chulia:2016ngi,  Ma:2015mjd, Reig:2016ewy, Bonilla:2016zef, Bonilla:2016diq, Ma:2016mwh, Ma:2017kgb, Borah:2016lrl, Borah:2016zbd, Borah:2016hqn, Borah:2017leo, CentellesChulia:2017koy, Bonilla:2017ekt,  Borah:2017dmk, CentellesChulia:2018gwr, CentellesChulia:2018bkz, Borah:2018gjk, Borah:2018nvu, CentellesChulia:2019xky,Jana:2019mgj, Borah:2019bdi, Dasgupta:2019rmf, Ma:2019byo, Ma:2019iwj, Saad:2019bqf, Jana:2019mez, Nanda:2019nqy, Chowdhury:2022jde, Narendra:2017uxl} and references therein. Here we consider a type II seesaw realisation of the dimension six operator for light Dirac neutrino mass \cite{CentellesChulia:2018bkz}. The particle content of the minimal model is shown in table \ref{tab1}. A vector-like fermion doublet $\Psi$ and a real scalar triplet $\Omega$ are introduced to realise the type II Dirac seesaw\footnote{This is not a unique choice involving a real scalar triplet. For example, one can consider vector-like fermion triplet instead of fermion doublet too to realise the same seesaw realisation of dimension six operator. Such choices do not change the analysis significantly and we stick to the minimal choice of fermion doublet.}. A softly broken discrete $Z_2$ symmetry is introduced to forbid direct coupling of right handed neutrino $\nu_R$ to the SM lepton and Higgs in the form of $\overline{L} \tilde{H} \nu_R$. An overall global lepton number symmetry is assumed as in conventional Dirac seesaw models to keep lepton number violating Majorana mass terms away from the Lagrangian.

	\begin{table}[h!]
		\small
		\begin{center}
			\begin{tabular}{||@{\hspace{0cm}}c@{\hspace{0cm}}|@{\hspace{0cm}}c@{\hspace{0cm}}|@{\hspace{0cm}}c@{\hspace{0cm}}|@{\hspace{0cm}}c@{\hspace{0cm}}||}
				\hline
				\hline
				\begin{tabular}{c}
					{\bf ~~~~ Gauge~~~~}\\
					{\bf ~~~~Group~~~~}\\ 
					\hline
					
					$SU(2)_{L}$\\ 
					\hline
					$U(1)_{Y}$\\ 
					\hline
					$Z_2$\\ 
				
				\end{tabular}
				&
				&
				\begin{tabular}{c|c|c}
					\multicolumn{3}{c}{\bf Fermion Fields}\\
					\hline
					~~~$L$ & $\Psi_{L, R}$~~~& ~~~$\nu_R$\\
					\hline
					$2$ & $2$&$1$\\
					\hline
					$-\frac{1}{2}$ & $-\frac{1}{2}$&$0$\\
					\hline
					$1$ & $-1$&$-1$ \\
				\end{tabular}
				&
				\begin{tabular}{c|c}
					\multicolumn{2}{c}{\bf Scalar Field}\\
					\hline
					~~~$\Omega$ & H\\
					\hline
					$3$ & 2\\
					\hline
					$0$ & $-\frac{1}{2}$ \\
					\hline
					$-1$ & 1 \\
		
				\end{tabular}\\
				\hline
				\hline
			\end{tabular}
			\caption{Relevant particles and their
				gauge charges in the tree level Dirac seesaw model.}
			\label{tab1}
		\end{center}    
	\end{table}
	
The relevant Yukawa Lagrangian can be written as
\begin{align}
	-\mathcal{L} & \supseteq M_\Psi \overline{\Psi} \Psi + y_L \overline{L} \Omega \Psi_R + y_{\nu_R} \overline{\Psi_L} \tilde{H} \nu_R+ {\rm h.c}	\label{lag} 
\end{align}
where $\Psi = (\psi^0, \psi^-)^T$ is the vector-like fermion doublet.

\begin{figure}[htb!]
    \centering
    \includegraphics[scale=0.5]{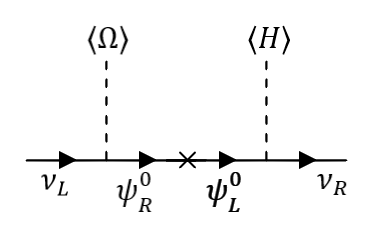}
    \caption{Origin of tree level Dirac Neutrino Mass}
    \label{fig1}
\end{figure}


Scalar potential of the minimal model can be written as 
\begin{align}
    V\left( H,\, \Omega\right) & = - \mu_{H}^2 H^\dagger H + \lambda_H \left(H^\dagger H \right)^2 - M_\Omega^2\, \text{Tr } \Omega^2 \nonumber \\
    & + \lambda_{\Omega}\, \text{Tr } \Omega^4 + \lambda_{\Omega}^\prime (\text{Tr } \Omega^2)^2 + \lambda_{H\Omega}\, \left(H^\dagger H\right) \text{Tr } \Omega^2  \nonumber \\ & +\lambda_{H\Omega}^\prime\, \left(H^\dagger\,\Omega^2\, H\right) + \zeta\, \left(H^\dagger\,\Omega\, H\right).
    \label{potential}
\end{align}
Here we consider $M_\Omega^2 <0$ such that the neutral component of $\Omega$ acquires only an induced vacuum expectation value (VEV) after electroweak symmetry breaking. After symmetry breaking, neutrinos will acquire a Dirac mass at tree level from the diagram shown in Fig. \ref{fig1}. This can be estimated as
\begin{eqnarray}
    m_{\nu}= \frac{y_L y_{\nu_R}\,\left<H\right>\,\left<\Omega \right>}{M_{\Psi}}.
\end{eqnarray}

\vspace{0.2cm}
\noindent
\textbf{\emph{W boson mass}:} Any new physics which could be responsible for the W boson mass anomaly, can be parametrised by oblique parameters S, T, U \cite{Peskin:1990zt, Peskin:1991sw}. Considering the U parameter to be suppressed, one can parametrise any BSM physics contribution to W boson mass in terms of $S, T$ parameters. Taking $\alpha_{\rm EW}, G_F, m_Z$ as input parameters, the required fitting of $S, T$ parameters in view of the recent $W$-mass anomaly has been discussed in \cite{Bagnaschi:2022whn}. It should be noted that, change in $S, T$ parameters due to BSM physics will also change the weak mixing angle $\theta_W$, which is precisely measured. The mass of W bson $m_W$ and $\sin^2_{\theta_W}(m_Z)_{\overline{MS}}$ can be expressed in terms of electroweak precision parameters namely, S and T as \cite{Kumar:2013yoa},
\begin{align}
    m_{W}&= 80.357\text{ GeV} \left(1-0.0036\,S+0.0056\,T \right), \nonumber \\
\sin^2_{\theta_W}(m_Z)_{\overline{\rm MS}}&= 0.23124 \left(1-0.0157\,S+0.0112\,T \right),  
\label{TS}
\end{align}
which implies that the compatibility of the enhanced W boson mass with precision measurements of $\theta_W$ require both S and T parameters to be non-zero, also seen from the fits presented in \cite{Bagnaschi:2022whn}. However, in the case of real scalar triplet, we will not get any correction to S parameter, even at one-loop level \cite{Forshaw:2001xq}. On the other hand, T parameter can get contribution from the induced VEV of the neutral component of triplet. The required limit on T, in order to explain the enhanced W boson mass, will be 
\begin{eqnarray}
\label{tpara}
T = 0.17\pm 0.020889.
\end{eqnarray}
However, according to Eq. \eqref{TS}, any change in T parameter with S=0 will put $\sin^2{\theta_W}$ in tension with the LEP data. The above mentioned range of T would imply that $\sin^2{\theta_W}$ should lie in between $0.230746-230854$. Additional contribution to $S, T$ parameters should be able to reduce this tension, as we will discuss later in the radiative version of Dirac seesaw.

As mentioned earlier, the presence of the trilinear coupling $\zeta\, \left(H^\dagger\,\Omega\, H\right) $ in equation \eqref{potential}, leads to an induced VEV of the neutral component of $\Omega$ given by
\begin{equation}
    v_{\Omega} = \frac{\sqrt{2} \zeta\,v^2}{4 \,M_{\Omega}^2 + 2 \left( \lambda_{H\Omega}+\lambda_{H\Omega}^\prime\right) v^2},
\end{equation}
where $v$ denotes the SM Higgs VEV. In the limit $(\lambda_{H\Omega} + \lambda_{H\Omega}^\prime)v^2\, \ll \, 2M_{\Omega}^2$, we can approximate the induced VEV as,
\begin{equation}
    v_{\Omega} = \frac{\zeta\,v^2}{2\sqrt{2} \,M_{\Omega}^2}.
    \label{veveq}
\end{equation}
Since $\Omega$ is a real scalar triplet with hypercharge zero, it will contribute only to W mass and not to Z mass at tree level. Therefore, it will change only the T parameter through its contribution to W-mass and can be expressed as,
\begin{eqnarray}
    T=\frac{v^2}{\alpha} \frac{\zeta^2}{M_{\Omega}^4},
\end{eqnarray}
where $\alpha$ is the the fine structure constant. It should be noted that, the vector-like fermion doublet $\Psi$, in spite of having $SU(2)_L$ gauge interactions, can not give rise to the required enhancement via radiative corrections due to degenerate masses of its components.

\vspace{0.2cm}

\begin{figure}[htb!]
    \centering
    \includegraphics[scale=0.35]{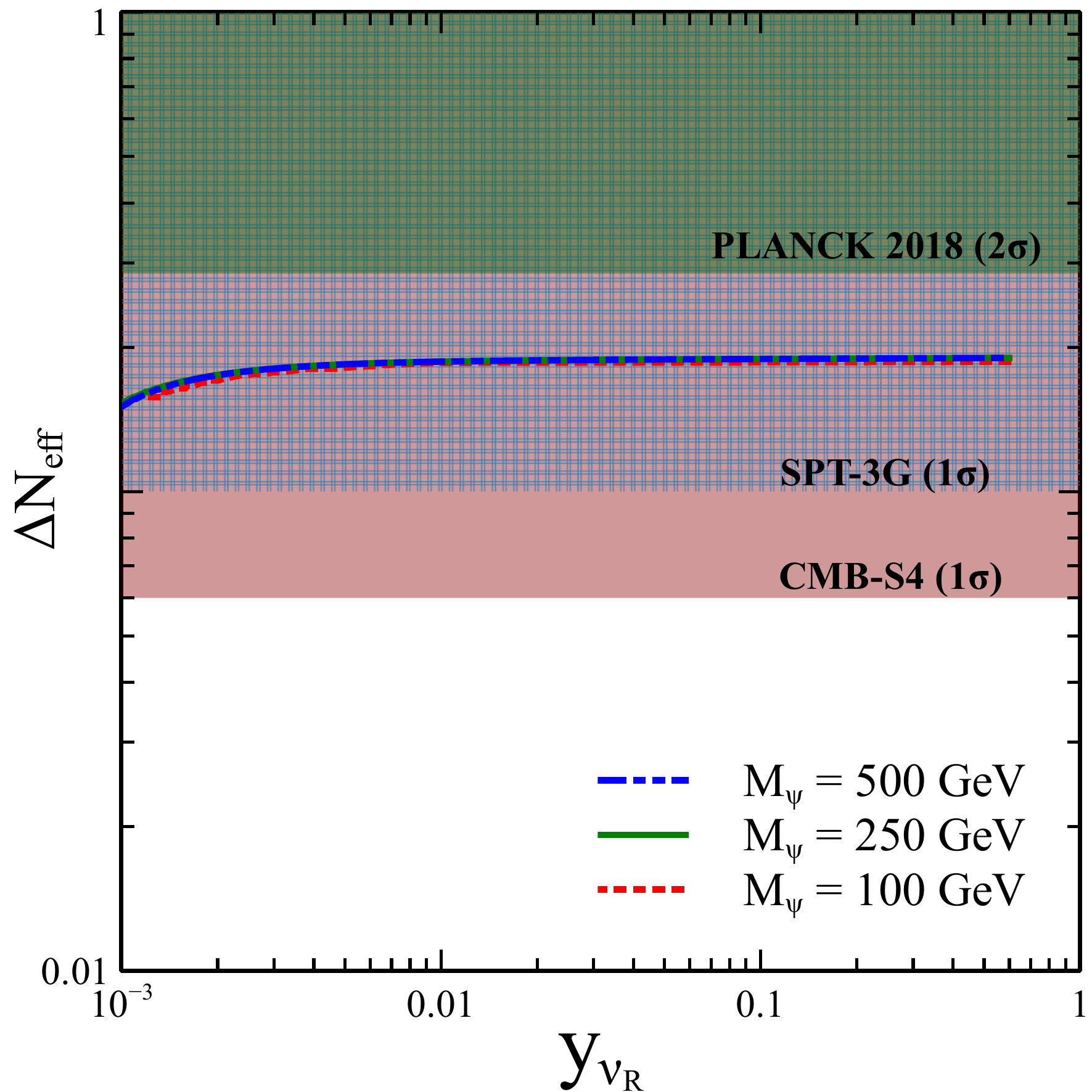}
    \caption{Contribution to $\Delta{N
    _{\rm eff}}$ as a function of the Yukawa interaction $y_{\nu_R}$ for three benchmark values of $M_\psi$.} 
    \label{fig:neff}
\end{figure}

\noindent
\textbf{\emph{Observable $\Delta N_{\rm eff}$}:}
Thermalisation of the $\nu_R$s gives extra contribution to the total radiation energy density of the universe $N_{\rm eff}$. The quantity $N_{\rm eff}$ is defined
as the contribution of non-photon components to the radiation energy
density normalised by the contribution of a single active neutrino
species ($\varrho_{\nu_L}$) \cite{Mangano:2005cc}
i.e.
\begin{eqnarray}
N_{\rm eff} &\equiv& \dfrac{\varrho_{\rm rad}-\varrho_{\gamma}}
{\varrho_{\nu_L}}\, ,
\label{def_neff}
\end{eqnarray}
which, during the
era of recombination ($z\sim 1100$), is restricted by Planck 2018 data \cite{Aghanim:2018eyx} to be
\begin{eqnarray}
{\rm
N_{eff}= 2.99^{+0.34}_{-0.33}
}
\label{Neff}
\end{eqnarray}
at $2\sigma$ CL including baryon acoustic oscillation (BAO) data.
At $1\sigma$ CL it becomes more stringent to $N_{\rm eff} = 2.99 \pm 0.17$.
Both these bounds are consistent with the standard model (SM)
prediction $N^{\rm SM}_{\rm eff}=3.045$ \cite{Mangano:2005cc, Grohs:2015tfy, deSalas:2016ztq}. Upcoming cosmic microwave background (CMB) experiments like SPT-3G \cite{Avva:2019hzz} and
CMB-S4 \cite{Abazajian:2016yjj} will be able to constrain $N_{\rm eff}$ upto a much higher accuracy, having the potential to verify BSM scenarios like ours which lead to enhancement in $N_{\rm eff}$. Some recent studies on light Dirac neutrinos and enhanced $\Delta N_{\rm eff}$ can be found in \cite{Abazajian:2019oqj, FileviezPerez:2019cyn, Nanda:2019nqy, Han:2020oet, Luo:2020sho, Borah:2020boy, Adshead:2020ekg, Luo:2020fdt, Mahanta:2021plx, Du:2021idh, Biswas:2021kio, Li:2022yna}. For thermalised $\nu_R$, the enhancement to $N_{\rm eff}$ can be expressed as,
\begin{equation}
    \Delta N_{\rm eff} = N_{\rm eff} - N_{\rm eff}^{\rm SM}=N_{\nu_R} \left( \frac{T_{\nu_R}}{T_{\nu_L}}\right)^4 = N_{\nu_R} \left( \frac{g_{*}\left(T_{\nu_L}^{\rm dec}\right)}{g_{*}\left(T_{\nu_R}^{\rm dec}\right)}\right)^{4/3}.
\end{equation}
This can be computed simply by assuming instantaneous decoupling of $\nu_R$'s and finding the corresponding decoupling temperature. This also agrees to a great accuracy with more detailed analysis involving Boltzmann equations \cite{Luo:2020sho, Biswas:2021kio}. The corresponding results are shown in Fig. \ref{fig:neff} for benchmark choices of $\Psi$ mass and Yukawa couplings which dictate the thermalisation as well as decoupling temperature of $\nu_R$. Clearly, the entire parameter space remains within the reach of future CMB experiments.
\begin{figure}[htb!]
    \centering
    \includegraphics[scale=0.35]{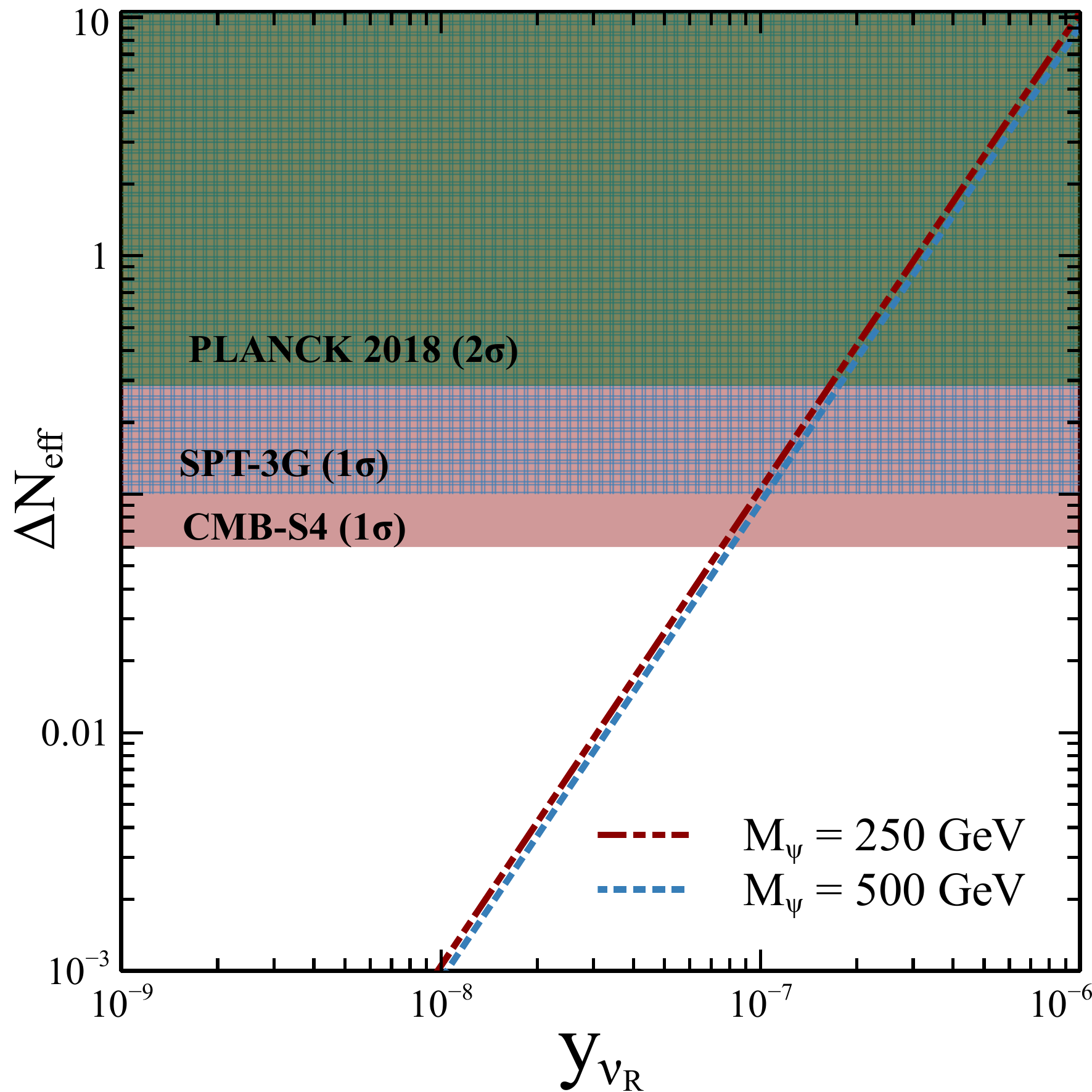}
    \caption{Non-thermal contribution to $\Delta{N
    _{\rm eff}}$ from the decay of fermion doublet as a function of the Yukawa interaction $y_{\nu_R}$.} 
    \label{fig:neff_nt}
\end{figure}
\begin{figure}
    \centering
    \includegraphics[scale=0.35]{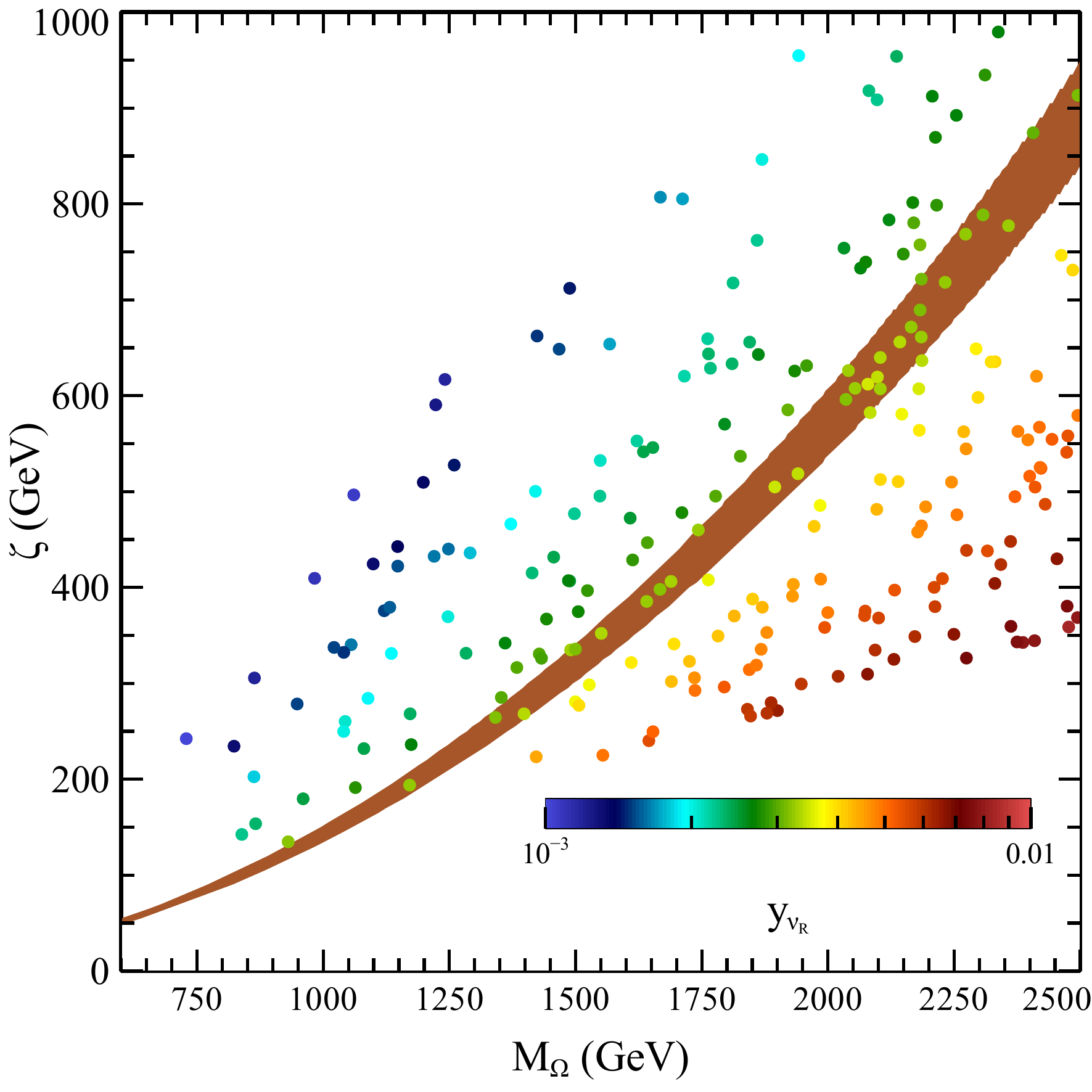}
    \caption{Allowed parameter space in $\zeta$ vs $M_{\Omega}$ plane from the T parameter constraint shown in equation \eqref{tpara}. Here we have fixed $M_\psi$ at 250 GeV and $y_L$ at $10^{-8}$.}
    \label{fig:T}
\end{figure}
However, $\nu_R$ can also be produced through freeze-in mechanism if the corresponding interactions are not sufficient to keep them in the thermal plasma. For non-thermal production of $\nu_R$, we can calculate the enhancement in $N_{\rm eff}$ by solving the Boltzmann equation which tracks the non-thermal production of $\nu_R$ from other particles \cite{Luo:2020fdt}. The corresponding Boltzmann equation can be written, for our model, as
\begin{equation} \label{case1_nuR}
    \frac{d Y_{\nu_R}}{dx} =    \frac{2\beta(x)}{H s^{1/3} x} \left<E\Gamma\right> Y_{\psi}^{eq}.
\end{equation}
where $x$ is defined as $M_{\psi}/T$, s is the entropy density of the universe, and  $\beta(T)=\left(1+\frac{T dg_{s}/dT}{3g_{s}}\right)$. The 2 factor in the right hand side of the Boltzmann equation is due to the Dirac nature of $\nu_R$. The collision term $<E\Gamma>$ can be written as 
\begin{eqnarray}
    \left<E\Gamma\right> = \frac{y_{\nu_R}^2 (M_{\psi}^2-m_{h}^2)^2}{32\pi M_{\psi}^2}\left(1-\frac{m_{h}^2}{M_{\psi}^2} \right).
\end{eqnarray}
Here, we consider that the $\nu_R$ produced from the decay of $\psi$ which was in thermal equilibrium with the SM particles owing to there gauge interactions. One can write the corresponding enhancement to $N_{\rm eff}$ as
\begin{equation}
    \Delta{N_{\rm eff}}= 3 \frac{s^{4/3} Y_{\nu_R}}{\rho_{\nu_L}}\bigg|_{T=T_{\rm CMB}},
\end{equation}
where the numerical value of $Y_{\nu_R}$ at $T=T_{\rm CMB}$ is found by solving equation \eqref{case1_nuR}. The results are shown in Fig . \ref{fig:neff_nt} for chosen benchmark parameters. As expected, with the rise in Yukawa coupling, the non-thermal contribution to $\nu_R$ density and hence $N_{\rm eff}$ increases. As can be noticed while comparing with Fig. \ref{fig:neff}, the Yukawa coupling remains suppressed in this case for $\nu_R$ to be in non-thermal regime.

In Fig. \ref{fig:T}, we show the scan in the plane of $\zeta$ and $M_\Omega$ which satisfy enhanced W-mass criteria reported by the CDF collaboration. The required enhancement in W-mass constrains the VEV of $\Omega$ to lie in few GeV regime which can be translated into constraints in $\zeta, M_{\Omega}$ plane following Eq.\eqref{veveq} as shown in Fig~\ref{fig:T}. In colour code, we show the Yukawa coupling strength of $\nu_R$ considering it to be thermally generated providing a large contribution to $N_{\rm eff}$ within reach of future CMB experiments, as shown in figure \ref{fig:neff}. We also take into account the constraint on the scale of neutrino mass. To generate the correct neutrino mass, the product of $y_L\, y_{\nu_R}$ have to be $\mathcal{O}(10^{-10})$ assuming $M_\psi$ and $\langle H \rangle =v$ have similar order. In case of very small $y_L$ (say $y_L\, \sim \,10^{-8}$), $y_{\nu_R}$ has to be large enough to generate the correct neutrino mass. Such interactions can lead to the thermalisation of $\nu_R$ in the early universe whose effects can be seen as the extra radiation energy density at the time of recombination. On the other hand, if we consider sizable $y_L$ this makes $y_{\nu_R}$ automatically small from the requirement of correct neutrino mass. In this case, although $\nu_R$ can not be thermalised due to such feeble interactions, they can still be produced from the non-thermal decay of $\psi$ as they were present in the thermal plasma due to their gauge interactions.

\begin{figure}
    \centering
    \includegraphics[scale=0.5]{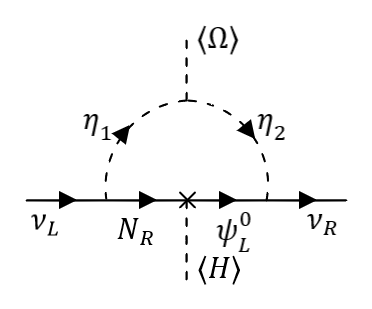}
    \caption{Origin of Scotogenic Dirac Neutrino Mass.}
    \label{fig3}
\end{figure}

\vspace{0.2cm}
\noindent
\textbf{\emph{Radiative Dirac Seesaw}:} The particle content of the model is shown in table \ref{tab2}. The relevant Lagrangian can be written as:
\begin{align}
	-\mathcal{L} & \supseteq M_\Psi \overline{\Psi} \Psi+ M_N \overline{N} N +Y \overline{\Psi}\widetilde{H}N + Y_{Nl} \overline{L} \tilde{\eta_1} N_R \nonumber \\&+ Y_R \overline{\Psi_L} \tilde{\eta_2} \nu_R 
 	+Y_{\psi l} \overline{\Psi_L} \eta_1 l_R+ {\rm h.c}	
 	\label{lag2}
\end{align}
One of the discrete symmetries namely $Z'_2$ will be broken while the other will remain unbroken ensuring the stability of dark matter. After the electroweak symmetry breaking, light Dirac neutrino masses arise at one-loop level as shown in Fig. \ref{fig3}. The corresponding one-loop contribution can be estimated following \cite{Farzan:2012sa, Borah:2016zbd}. The analysis for W boson mass correction from triplet VEV and enhancement of $N_{\rm eff}$ from thermal or non-thermal $\nu_R$ remain more or less same as in the minimal model and hence we only comment on additional phenomenology of the radiative model below. It should be noted that similar to the tree level Dirac seesaw model discussed earlier, here also we consider a global lepton number symmetry in order to keep the Majorana mass terms away.

	\begin{table}[h!]
		\small
		\begin{center}
			\begin{tabular}{||@{\hspace{0cm}}c@{\hspace{0cm}}|@{\hspace{0cm}}c@{\hspace{0cm}}|@{\hspace{0cm}}c@{\hspace{0cm}}|@{\hspace{0cm}}c@{\hspace{0cm}}||}
				\hline
				\hline
				\begin{tabular}{c}
					{\bf ~~~~ Gauge~~~~}\\
					{\bf ~~~~Group~~~~}\\ 
					\hline
					
					$SU(2)_{L}$\\ 
					\hline
					$U(1)_{Y}$\\ 
					\hline
					$Z_2$\\ 
					\hline 
					$Z'_2$
				
				\end{tabular}
				&
				&
				\begin{tabular}{c|c|c}
					\multicolumn{3}{c}{\bf Fermion Fields}\\
					\hline
					~~~$N_{L,R}$ & $\Psi_{L, R}$~~~& ~~~$\nu_R$\\
					\hline
					$1$ & $2$&$1$\\
					\hline
					$0$ & $-\frac{1}{2}$&$0$\\
					\hline
					$-1$ & $-1$&$1$ \\
					\hline
					$1$ & $1$&$-1$ \\
				\end{tabular}
				&
				\begin{tabular}{c|c|c}
					\multicolumn{3}{c}{\bf Scalar Field}\\
					\hline
					~~~$\Omega$ & $\eta_1$ & $\eta_2$\\
					\hline
					$3$ & 2 & 2\\
					\hline
					$0$ & $-\frac{1}{2}$ & $-\frac{1}{2}$ \\
					\hline
					$1$ & -1 & -1\\
					\hline
					$-1$ & 1 & -1 \\
		
				\end{tabular}\\
				\hline
				\hline
			\end{tabular}
			\caption{New Particles and their
				gauge charges in the radiative Dirac seesaw model.}
			\label{tab2}
		\end{center}    
	\end{table}

After electroweak phase transition, the VEV of SM Higgs leads to mixing between the neutral component of doublet $\psi^0$ (as ($\Psi^T=(\psi^0, \psi^-)$) and $N$ giving rise to the well motivated singlet-doublet DM Model which has been studied extensively in the literature \cite{Mahbubani:2005pt,DEramo:2007anh,Enberg:2007rp,Cohen:2011ec,Cheung:2013dua, Restrepo:2015ura,Calibbi:2015nha,Cynolter:2015sua, Bhattacharya:2015qpa,Bhattacharya:2017sml, Bhattacharya:2018fus,Bhattacharya:2018cgx,DuttaBanik:2018emv,Barman:2019tuo,Bhattacharya:2016rqj, Calibbi:2018fqf, Barman:2019aku, Dutta:2020xwn, Borah:2021khc, Borah:2021rbx, Borah:2022zim}. As clear from the Lagrangian in Eq.~\eqref{lag2}, here we consider a Dirac version of the singlet-doublet DM model. The mass terms for these fields can then be written together as follows:
\begin{eqnarray}
-\mathcal{L}^{\rm VF}_{\rm mass}&=&M_\Psi\overline{\psi^0}\psi^0+M_\Psi{\psi^+}\psi^-+M_{N} \overline{N}N \nonumber\\&+&  \frac{Y v}{\sqrt2}~ \overline{\psi^0}~N 
+\frac{Y v}{\sqrt2} ~\overline{N}~\psi^0 \nonumber \\
&=&
\overline{\left(\begin{matrix}
N & \psi^0 
\end{matrix}\right)}
{\left(\begin{matrix}
M_{N} & Y v/\sqrt2\\
Y v/\sqrt2 & M_\Psi
\end{matrix}\right)}
{\left(\begin{matrix}
N \\ \psi^0
\end{matrix}\right)}\nonumber\\
&+&M_\Psi {\psi^+}\psi^- . \nonumber\\
\end{eqnarray}

The unphysical basis, $\left(\begin{matrix}
 N &&  \psi^0 
\end{matrix}\right)^T$ is related to physical basis, $\left(\begin{matrix}
 \psi_1 &&  \psi_2 
\end{matrix}\right)^T$ through the following unitary transformation: 
\begin{eqnarray}
 \left(\begin{matrix}
 N \\ \psi^0
\end{matrix}\right) 
=\mathcal U \left(\begin{matrix}
\psi_1 \\  \psi_2 
\end{matrix}\right)=\left(\begin{matrix}
 \cos\theta & -\sin\theta \\
 \sin\theta & \cos\theta
 \end{matrix}\right) 
\left(\begin{matrix}
\psi_1 \\  \psi_2 
\end{matrix}\right),
\end{eqnarray}
where the mixing angle is given by:
\begin{eqnarray}\label{ref:mixang}
\tan{2\theta}= - \frac{\sqrt2 Y v}{M_\Psi-M_{N}} .
\end{eqnarray}
The mass eigenvalues of the physical states $\psi_1$ and $\psi_2$ are respectively given by:
\begin{eqnarray}\label{ref:phymass}
M_{\psi_1} &=& M_{N} \cos^2\theta + M_\Psi \sin^2\theta + \frac{Y v}{\sqrt2}\sin 2\theta \nonumber \\
M_{\psi_2} &=& M_{N} \sin^2\theta + M_\Psi \cos^2\theta - \frac{Y v}{\sqrt2}\sin 2\theta
\end{eqnarray}
For small $\sin\theta$ ($\sin\theta \rightarrow 0$) limit, $M_{\psi_1}$  and $M_{\psi_2}$  can be further expressed as:
\begin{eqnarray}
M_{\psi_1} &\simeq& M_{N}+\frac{Y v}{\sqrt 2}\sin{2\theta} \equiv  M_{N}-\frac{(Y v)^2}{(M_\Psi-M_{N})} , \nonumber \\ 
M_{\psi_2} &\simeq& M_{\Psi}-\frac{Y v}{\sqrt 2}\sin{2\theta} \equiv  M_{\Psi}+\frac{(Y v)^2}{(M_\Psi-M_{N})} .
\end{eqnarray}
As $Y v/\sqrt2 \ll M _{N} < M_\Psi$. Hence $M_{\psi_1} < M_{\psi_2}$ and thus $\psi_1$ becomes the stable DM candidate. From Eqs.\eqref{ref:mixang} and \eqref{ref:phymass}, one can write $Y$ and $M_{\psi}$ in terms of $M_{\psi_{1,2}}$ and $\sin\theta$ as :
\begin{eqnarray}\label{ref:reltn}
 Y &=& - \frac{\Delta{M} \sin{2\theta}}{\sqrt2 v}, \nonumber \\
 M_\Psi &=& M_{\psi_1}\sin^2\theta + M_{\psi_2} \cos^2\theta .
\end{eqnarray}
where $\Delta M = M_{\psi_2} - M_{\psi_1}$

Thus the interaction terms of the dark sector fermions in the mass basis is:
\begin{widetext}
\begin{eqnarray}
\mathcal{L}^{\rm DM}_{\rm int} &=& \overline{\Psi}~[i\gamma^{\mu}(\partial_{\mu} - i g \frac{\sigma^a}{2}W_{\mu}^a - i g^{\prime}\frac{Y}{2}B_{\mu})]~\Psi + \overline{N}~(i\gamma^\mu \partial_{\mu})~N - (Y \overline{\Psi}\widetilde{H}N+{\rm h.c.})\nonumber\\
\mathcal{L}^{\rm DM}_{\rm int} &=& g_Z \Big[\sin^2\theta \overline{\psi_1}\gamma^{\mu}Z_{\mu}\psi_1+\cos^2\theta \overline{\psi_2}\gamma^{\mu}Z_{\mu}\psi_2 
+\sin\theta \cos\theta(\overline{\psi_1}\gamma^{\mu}Z_{\mu}\psi_2+\overline{\psi_2}\gamma^{\mu}Z_{\mu}\psi_1)\Big]   \nonumber \\
&+&g_W(\sin\theta \overline{\psi_1}\gamma^\mu W_\mu^+ \psi^- + \cos\theta \overline{\psi_2}\gamma^\mu W_\mu^+ \psi^- )  +g_W( \sin\theta{\psi^+}\gamma^\mu W_\mu^- \psi_1 + \cos\theta {\psi^+}\gamma^\mu W_\mu^- \psi_2 ) \nonumber \\
&-& g_Z ~c_{2w}~ {\psi^+}\gamma^{\mu}Z_{\mu}\psi^- - e_0 {\psi^+}\gamma^{\mu}A_{\mu}\psi^- -\frac{Y}{\sqrt2}h\Big[\sin2\theta(\overline{\psi_1}\psi_1-\overline{\psi_2}\psi_2)+\cos2\theta(\overline{\psi_1}\psi_2+\overline{\psi_2}\psi_1)\Big]
\end{eqnarray}
\end{widetext}
where $g_Z=e_0/2 s_w c_w$, $g_W=e_0/\sqrt{2} s_w$ and $s_w$, $c_w, c_{2w}$ denote $\sin \theta_W$ and $\cos \theta_W$ and $\cos2\theta_{W}$ respectively.


The important parameters which decide the relic abundance of $\psi_1$ are mass of DM ($M_{\psi_1}$), the 
mass splitting ($\Delta M$) between the DM and the next-to-lightest stable particle (NLSP) and the singlet-doublet mixing angle $\sin \theta$. Here we have used \texttt{micrOmega}~\cite{Belanger:2008sj} to calculate the relic density of DM.
In this scenario the annihilation cross-section increases with $\sin\theta$, as it enhances the $SU(2)$ component and hence results in smaller relic density.  When the mass-splitting is small it leads to effective co-annihilation and hence reduces the relic density contribution(due to less Boltzmann suppression). In this case the effect of $\sin \theta$ on relic abundance is quite negligible whereas for large $\Delta M$, as the co-annihilation becomes suppressed, 
the effect of $\sin \theta$ or relic abundance is clearly visible. For small $\sin \theta$, the effective annihilation cross-section is small leading 
to a large relic abundance, while for large $\sin \theta$ the relic abundance is small provided that the $\Delta M$ is large enough to reduce the co-annihilation contributions.
\begin{figure}[!htb]
\includegraphics[height=6.5cm,width=7.5cm]{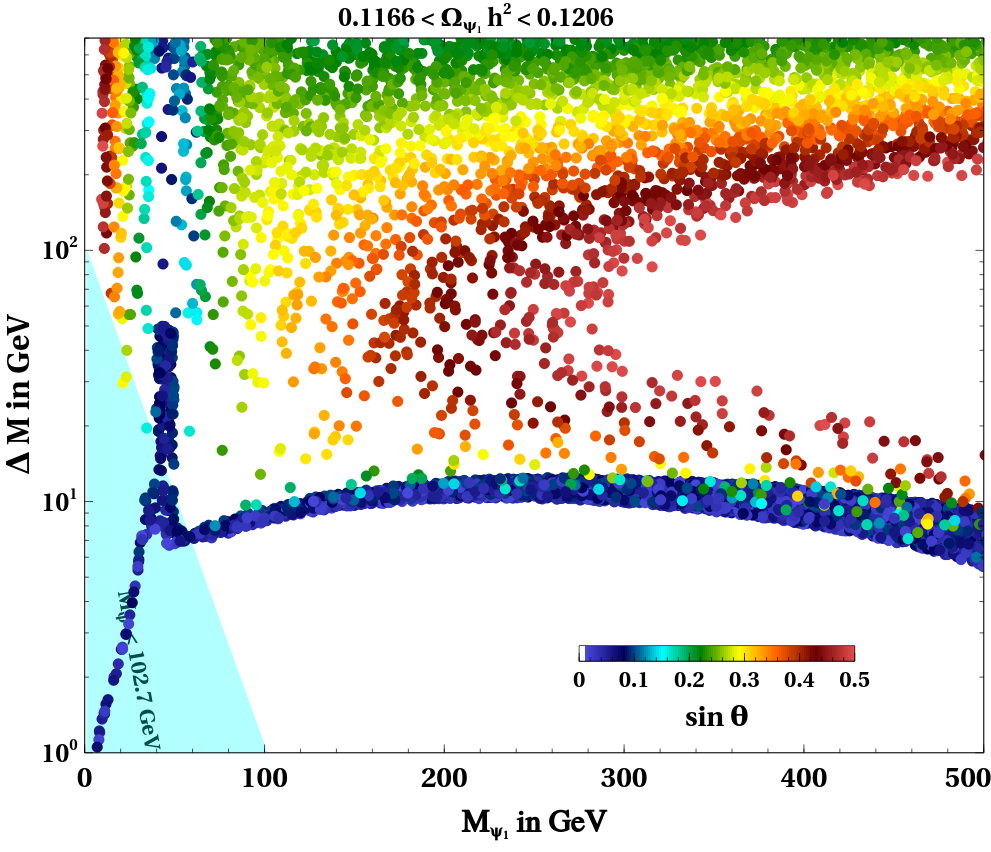}
\caption{DM Relic density (from PLANCK) allowed parameter space shown in the plane of $M_{\psi_{1}}~ vs~ \Delta M$}\label{relic}
\end{figure}

In Fig.~\ref{relic}, we have shown the points satisfying correct relic density in the plane of $M_{\psi_1}$ versus $\Delta M$. We can see that, for a wide range of singlet-doublet mixing ($\sin \theta $) which is indicated by the color bar, we can get correct relic abundance.
This result can be explained by understanding the interplay between $\sin\theta$ and $\Delta M$ in deciding the effective annihilation cross-section of the DM. If we divide the plane of $M_{\psi_1}$ versus $\Delta m$ into two regions: (I) the bottom portion with small $\Delta m$, where
$\Delta M$ decreases with larger mass of $\psi_1$, (II) the top portion with larger mass splitting $\Delta m$, where $\Delta m$ increases slowly 
with larger DM mass $M_{\psi_1}$. In the former case, for a given range of $\sin \theta$, the annihilation cross-section decreases as mass of DM increases. Therefore, more co-annihilation contribution is needed for compensation, which requires $\Delta M$ to decrease. This also implies that the region 
below this is under abundant as small mass-splitting imply large co-annihilation, while the region above, is over abundant as large $\Delta M$ lead to small co-annihilation for a given mass of DM. Now in region 
(II), as $\Delta M$ is large, co-annihilation contribution is much smaller here and hence the annihilation processes effectively decides the relic density. However as the Yukawa coupling $Y \propto \Delta M \sin\theta$, for a given $\sin \theta$, larger 
$\Delta M$ leads to larger $Y$ and therefore larger annihilation cross-section leading to under abundance, which can only be brought to correct ballpark by having a larger DM mass. That is the reason, in case-(II), the region above the shown allowed region of 
correct relic density is under abundant, while the region below it is over abundant.
\begin{figure}[!htb]
\includegraphics[height=6.5cm,width=7.5cm]{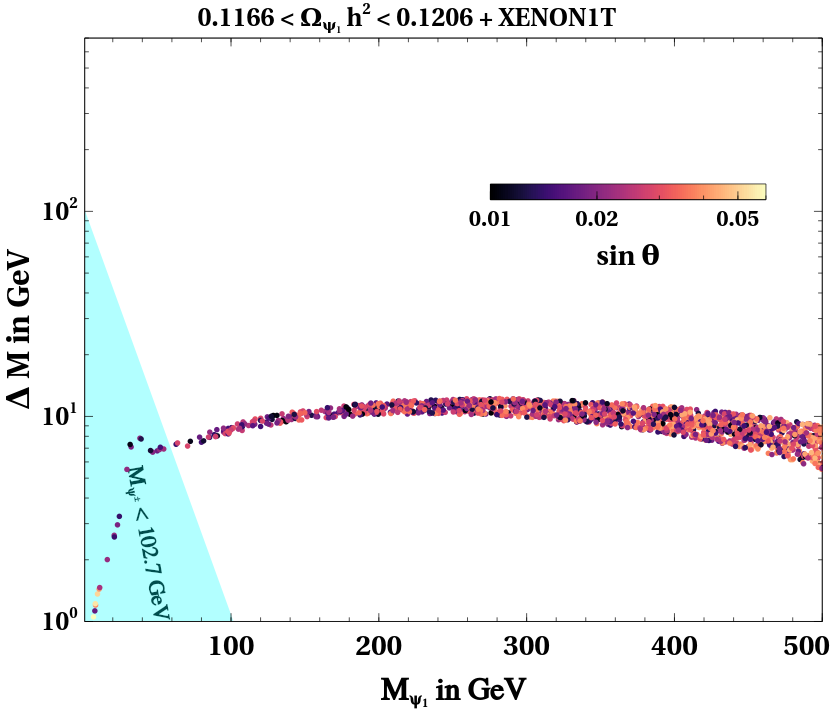}
\caption{DM Relic density (from PLANCK) + Direct search (from XENON1T) allowed parameter space shown in the plane of $M_{\psi_{1}}~ vs~ \Delta M$}\label{relicddsum}
\end{figure}

Now imposing the constraints from DM direct search experiments on top of the relic density allowed parameter space (Fig.~\ref{relic}) in the $\Delta M$ versus $M_{\psi_1}$ plane, we obtain Fig.~\ref{relicddsum}, which shows that the parameter space is crucially tamed down as compared to the parameter space satisfying correct relic density. As in this scenario,
due to the singlet-doublet 
mixing, the DM $\psi_1$ can scatter off the target nucleus at terrestrial direct search experiments, via Z and Higgs mediated processes, the cross-section for $Z$-boson mediated DM-nucleon scattering is given 
by~\cite{ Goodman:1984dc,Essig:2007az}
\begin{equation}\label{DM-nucleon-Z}
\sigma_{\rm SI}^Z = \frac{G^2_F \sin^4\theta}{\pi A^2 }\mu_r^2 \Big|\left[ Z f_p + (A-Z)f_n \right]^2\Big|^2   
\end{equation}
and similarly the spin-independent 
DM-nucleon cross-section through Higgs mediation is given by
\begin{equation}
		\label{dda2}
		\begin{aligned}
			\sigma^h_{\rm SI} &= \frac{4}{\pi A^2}\mu^2_r\frac{Y^2 \sin^2 2\theta}{M^4_h}\Big[\frac{m_p}{v}\Big(f^{p}_{Tu} + f^{p}_{Td} + f^{p}_{Ts} + \frac{2}{9}f^{p}_{TG}\\
			&+\frac{m_n}{v}\Big(f^{n}_{Tu} + f^{n}_{Td} + f^{n}_{Ts} + \frac{2}{9}f^{n}_{TG}\Big)\Big]^2
		\end{aligned}
	\end{equation}
It clearly shows that if $\sin \theta$ is large, then the the interaction strength is large and hence the DM-nucleon cross-section becomes large. Thus the direct search experiments constraints $\sin \theta$ to a great extent. This stringent upper limit on $\sin\theta$ is $\sin\theta \le 0.05$. As a result,only small $\Delta M$ region becomes viable, as correct relic density can be achieved dominantly through co-annihilation processes.

\begin{figure}[htb!]
    \centering
    \includegraphics[scale=0.5]{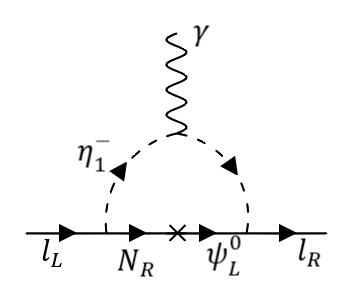}
    \caption{Lepton $(g-2)$ in scotogenic Dirac model.}
    \label{fig4}
\end{figure}

Recently, the the E989 experiment at Fermilab has measured the muon anomalous magnetic moment $a_\mu$ = $(g - 2)_\mu/2$ reporting a 4.2 $\sigma$ observed excess of $\Delta a_\mu = 251(59) \times 10^{-11}$ \cite{Muong-2:2021ojo}. On the other hand, the measurement of the fine structure constant using Cesium atoms \cite{Parker:2018vye} has led to a discrepancy in electron anomalous magnetic moment $a_e$ in negative direction with $2.4\sigma$ statistical significance.
In our model, the one-loop contribution to lepton anomalous magnetic moment $(g-2)_l$ can arise with the $Z_2$-odd particles in the loop as shown in Fig. \ref{fig4}. This contribution is given by~\cite{Calibbi:2018rzv, Jana:2020joi}:
\begin{eqnarray}
		\Delta a_l &=& -\frac{m_l }{16\pi^2 M^2_{\eta^+}}\sin 2\theta (Y_{N l} Y_{\psi l}) \nonumber\\&\times&\Bigg[M_{\psi_{_1}}F\bigg(\frac{M^2_{\psi_{1}}}{M^2_{\eta^+}}\bigg)-M_{\psi_{_2}}F\big(\frac{M^2_{\psi_{2}}}{M^2_{\eta^+}}\big)\Bigg]\nonumber\\
		&-&\frac{m^2_l }{8\pi^2 M^2_{\eta^+}}(Y^2_{N l}\cos^2\theta + Y^2_{\psi l}\sin^2\theta)G\bigg(\frac{M^2_{\psi_{_1}}}{M^2_{\eta^+}}\bigg)\nonumber\\ &-&\frac{m^2_l }{8\pi^2 M^2_{\eta^+}}(Y^2_{\psi l}\cos^2\theta + Y^2_{N l}\sin^2\theta)G\bigg(\frac{M^2_{\psi_{_2}}}{M^2_{\eta^+}}\bigg)\nonumber\\
	\end{eqnarray}
Where the loop functions $F$ and $G$ are given by
\begin{eqnarray}
F(x)&=&\frac{1-x^2+2x\log{x}}{2 (1-x)^3}\nonumber\\
G(x)&=&\frac{1-6x+3x^2+2x^3-6x^2\log x}{12(1-x)^4}
\end{eqnarray}
\begin{figure}
    \centering
    \includegraphics[scale=0.5]{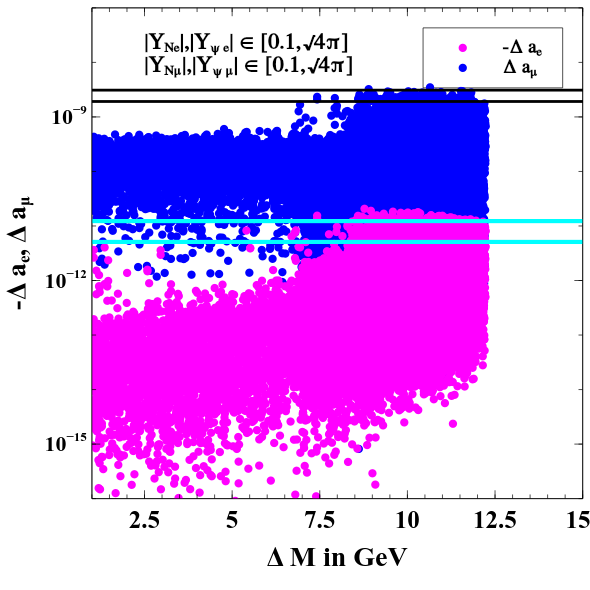}
    \caption{Lepton $(g-2)$ in radiative Dirac seesaw model.}
    \label{g2chk}
\end{figure}
We calculated the anomalous magnetic moment for both electron and muon with the parameter space consistent with relic and direct search constraints for DM and the result is shown in Figure~\ref{g2chk}. We find that this radiative Dirac seesaw model with singlet-doublet DM can give rise to both electron and muon $(g-2)$ in certain region of parameter space only when the relevant couplings $Y_{\psi l}$ and $Y_{N l}$ are large approaching the perturbativity limit. The black and cyan lines show the correct ball park for $\Delta a_{\mu}$ and $\Delta a_{e}$ respectively. It is worth mentioning here that the relative sign of $\Delta a_e$ and $\Delta a_\mu$ can be easily achieved by appropriately choosing the sign of the relevant Yukawa couplings.

Here it is worth mentioning that the presence of the inert scalar doublets in the radiative Dirac seesaw model can also contribute to the correction to W boson mass. However if the mass-splitting between the charged and the neutral components of the doublet is within 10 GeV and if the mass of the doublet is chosen beyond $800$ GeV or so, then it can not explain the W-mass anomaly as in such a case the S and T parameters are found to be of the order $\mathcal{O}(10^{-5})$ and $\mathcal{O}(10^{-3})$ respectively. Similar conclusion was also given in\cite{Han:2022juu}. For simplicity, we have considered a parameter space for the inert doublet scalar such that its co-annihilations with the DM are also negligible implying that inert doublet components remain much heavier compared to DM. Given that DM mass lies upto a few hundred GeV and inert doublet mass remains heavier, the dominant correction to W boson mass in our setup is coming from the triplet scalar VEV only. In addition, the presence of the additional singlet-doublet fermion also does not affect the S and T parameter significantly to explain the CDF-II results as the mass-splitting between the charged and the neutral components remain extremely small as a consequence of the upper limit on mixing angle $\sin \theta$ from DM direct search experiments\footnote{Similar conclusions are obtained even for Majorana singlet-doublet DM \cite{Borah:2022zim}.}. Thus, the singlet-doublet Dirac fermion DM with mass upto a few hundred GeV with desired relic and direct detection cross section automatically keeps the contribution of $Z_2$-odd particle's radiative contribution to W-mass suppressed. We find that after imposition of the relevant DM constraints, the allowed parameter space can give rise to S and T parameter of the order $\mathcal{O}(10^{-4})$ and $\mathcal{O}(10^{-7})$ respectively with the $Z_2$-odd fermions in the loop. If more generations of singlet-doublet fermion is introduced to generate non-zero neutrino mass to all three SM neutrinos, they can be kept much heavier so that their contribution to DM, $g-2$ phenomenology as well as W boson mass correction remain negligible.

As pointed out earlier while discussing the W boson mass in tree level Dirac seesaw model, it is in fact desirable if additional fields can also contribute to W-mass via contribution to both S, T parameters. While tree level model gives $S=0, T \neq 0$, we can have both S, T non-zero in the radiative model. However, if we demand that the dominant T contribution comes only from triplet VEV, then suppressing radiative contribution of inert scalar doublets to T also suppresses their contribution to S to some extent. We keep the mass splitting between charged and neutral components of inert scalar doublets below 20 GeV such that their contribution to T parameter at one loop remain small $T_\eta < 0.01$. In such a case, we could get their maximum contribution to S parameter to be approximately $S_\eta \leq 0.003$, which still keeps $\theta_W$ away from the LEP estimate. Considering the possibility of both scalar doublets and triplet via radiative and tree level contribution precision parameters respectively, should be able to reduce this tension.

\vspace{0.2cm}
\noindent
\textbf{\emph{Conclusion}:} We have proposed a Dirac seesaw mechanism for light neutrinos where a real scalar triplet plays a non-trivial role. In the minimal model with tree level seesaw, vector-like fermion doublets are also introduced to allow triplet couplings to leptons leading to the seesaw mechanism. The induced VEV of the neutral component of the triplet not only takes part in neutrino mass generation but also provides the necessary enhancement in W boson mass, in view of the recently reported results by the CDF collaboration of Fermilab. Depending upon triplet and vector-like fermion doublet couplings to right handed neutrinos, we find that additional effective relativistic degrees of freedom $N_{\rm eff}$ can be generated in the early universe either thermally or non-thermally. While Planck 2018 data already constrains such enhancement in $N_{\rm eff}$, future CMB experiments will be able to prove most part of the parameter space, providing a complementary probe of the model. We then discuss a radiative version of the same Dirac seesaw model after including additional scalar doublets and fermion singlets which also leads to the realisation of singlet-doublet Dirac fermion dark matter phenomenology. The model can also accommodate anomalous magnetic moments of electron and muon provided the respective Yukawa couplings are of order one in order to remain consistent with other phenomenological requirements. The proposed scenarios provide interesting phenomenology and complementary probes of real scalar triplet origin of W-mass anomaly in the context of tree level and radiative Dirac seesaw. Apart from the rich phenomenology provided by the radiative model, it also has the potential to reduce the tension between W mass and precision measurement of $\theta_W$ which exist in the pure triplet origin of W mass anomaly. Incorporating comparable contribution of scalar doublets at one loop and scalar triplet at tree level to W mass should be able to relax this condition. A detailed exploration of such a possibility along with the change in DM, g-2 results is left for future studies.

\noindent
\acknowledgments
 DN would like to thank Sanjoy Mandal for useful discussions. The work of DN is supported by National Research Foundation of Korea (NRF)’s grants, grants no. 2019R1A2C300500913. SM would like to thank Purusottam Ghosh for useful discussion. NS acknowledges the support from Department of Atomic Energy (DAE)- Board of Research in Nuclear Sciences (BRNS), Government of India (Ref. Number: 58/14/15/2021- BRNS/37220).

	\bibliographystyle{apsrev}
	\bibliography{ref, ref1.bib, ref5}


\end{document}